# *Multiple-relaxation-time lattice Boltzmann model for convection heat transfer in porous media under local thermal non-equilibrium condition*

Q. Liu, Y.-L. He

*Key Laboratory of Thermo-Fluid Science and Engineering of Ministry of Education, School of Energy and Power Engineering, Xi'an Jiaotong University, Xi'an, Shaanxi, 710049, China*

**Abstract**

In this paper, a multiple-relaxation-time (MRT) lattice Boltzmann (LB) model is proposed for convection heat transfer in porous media under local thermal non-equilibrium (LTNE) condition. The model is constructed within the framework of the three-distribution-function approach: two temperature-based MRT-LB equations are proposed for the temperature fields of fluid and solid phases in addition to the MRT-LB equation of a density distribution function for the velocity field described by the generalized non-Darcy model. The thermal non-equilibrium effects are incorporated into the model by adding source terms into the temperature-based MRT-LB equations. Moreover, the discrete lattice effects are considered in the introduction of source terms into the temperature-based MRT-LB equations. The source terms accounting for the thermal non-equilibrium effects are simple and the model retains the inherent features of the standard LB method. Numerical results demonstrate that the proposed model can be served as an accurate and efficient numerical method for studying convection heat transfer in porous media under LTNE condition.

## 1. Introduction

The analysis of convection heat transfer in porous media is of great importance due to its applications in many fields of science and engineering, such as geology, hydrology, geothermal energy systems, electronic device cooling, catalytic reactors, crude oil extraction and others. Consequently, convection heat transfer in porous media has been studied extensively by many researchers in the past several decades [1-3]. In the literature, two different models, i.e., the local thermal equilibrium (LTE) model and the local thermal non-equilibrium (LTNE) model, are used to describe convection heat transfer in porous media. The LTE model assumes that the local temperature of the fluid phase is equal to that of the solid phase, i.e., the interphase temperature difference can be neglected. LTE is an often-used assumption when simulating convection heat transfer in porous media. However, the assumption of LTE between the fluid and the solid phases is inadequate in many practical applications [4]. When modeling heat transfer process in porous media under a rapid heating or cooling condition, the fluid and solid phases are not at the same temperature at a local level [5]. When there is a significant heat generation occurring in any one of the two phases (either fluid or solid), the heat transfer process cannot be regarded as being in LTE state [6]. In addition, if the solid-to-fluid thermal conductivity ratio or heat capacity ratio is very large, the assumption of LTE must be discarded [7]. Therefore, the LTNE model must be employed to take account of the interphase temperature difference. In the past two decades, research into convection heat transfer in porous media under LTNE condition has received considerable interest, and various conventional numerical methods (e.g., FDM, FVM, FEM) have been used to study the thermal non-equilibrium effects of convection heat transfer in porous media (see Ref. [3] and references therein).

The lattice Boltzmann (LB) method [8-11], as a mesoscopic numerical method originated from the lattice-gas automata (LGA) method [12], has achieved great success in simulating fluid flows and modeling physics in fluids [13-20]. Unlike conventional numerical methods based on a direct discretization of the macroscopic continuum equations, the LB method is based on the mesoscopic kinetic equation for single-particle distribution function. Owing to its kinetic background, the LB method has some attractive advantages over the conventional numerical methods [21]: (i) non-linearity (collision) is local and non-locality (streaming) is linear, whereas the transport term $\mathbf{u} \cdot \nabla \mathbf{u}$ in the Navier-Stokes (N-S) equations is non-linear and non-local at a time; (ii) the pressure of the LB method is simply calculated by an equation of state, while in conventional numerical methods it is usually necessary and costly to solve a Poisson equation for the pressure field of the incompressible N-S equations; (iii) complex boundary conditions can be easily formulated in terms of the elementary mechanical rules; (iv) nearly ideal for parallel computing with very low communication/computation ratio.

The LB method has also been successfully applied to study fluid flow and convection heat transfer in porous media. In fact, the LB method was already applied to study porous flows by Succi et al. in 1989 [22]. The existing LB models for fluid flow and convection heat transfer in porous media can be generally classified into two categories, i.e., the pore scale method [22-26] and the representative elementary volume (REV) scale method [27-36]. In the pore scale method, fluid flow and heat transfer in the pores of the medium is directly modeled by the standard LB method, and the interaction between the fluid and solid phases is realized by using the no-slip bounce-back scheme. The detailed flow information of the pores can be obtained by this method, which can be utilized to investigate macroscopic relations. In the REV scale method, an additional term is incorporated into the LB equation to account for the presence of a porous medium based on some semi-empirical models (e.g.,

Darcy model [28], Brinkman-extended Darcy model [27], and generalized non-Darcy model [29-36]). After nearly two decades of development, the REV scale method has been developed into a numerically accurate and computationally efficient numerical tool for studying fluid flow and convection heat transfer in porous media on large scales. It is noted that the above mentioned REV-LB models [30-36] for convection heat transfer in porous media are based on the assumption of LTE.

Very recently, Gao et al. [37] proposed a thermal Bhatnagar-Gross-Krook (BGK) LB model for convection heat transfer in porous media under LTNE condition. Gao et al.'s model utilizes three different LB equations: a density evolution equation for the velocity field, and two temperature evolution equations for the temperature fields of fluid and solid phases. Numerical results indicate that Gao et al.'s model is suitable and efficient to study convection heat transfer in porous media under LTNE condition. However, the source terms accounting for the thermal non-equilibrium effects in Gao et al.'s model contain time derivative terms, thus finite difference scheme is needed in computations. The time derivative terms introduce additional complexity into the model and may do harm to the numerical stability. We would like to point out that the source terms accounting for the thermal non-equilibrium effects are actually in linear relationship with the temperatures of fluid and solid matrix, therefore, the source terms can be directly added into the temperature evolution equations by using appropriate scheme without discrete lattice effects.

Hence, in this paper, we aim to present a thermal multiple-relaxation-time (MRT) LB model for convection heat transfer in porous media under LTNE condition. In the model, the source terms accounting for the thermal non-equilibrium effects are treated in a simple way without discrete lattice effects. The rest of this paper is organized as follows. The macroscopic governing equations are briefly described in Section 2. In Section 3, the MRT-LB model for convection heat transfer in porous media under LTNE condition is presented in detail. The numerical results and some discussions are given in Section 4. Finally, some conclusions are made in Section 5.

## 2. Macroscopic governing equations

For convection heat transfer in an isotropic, rigid and homogeneous porous medium at the REV scale, assuming that the fluid is a normal Boussinesq fluid and the flow is incompressible without viscous heat dissipation and dispersion effect. In addition, the thermophysical properties of the fluid and solid phases are assumed to be constant over the range of temperatures considered. Based on the generalized non-Darcy model [38], the macroscopic governing equations for convection heat transfer in porous media under LTNE condition can be written as follows [3]:

Continuity equation

$$\nabla \cdot \mathbf{u} = 0, \qquad (1)$$

Momentum equation

$$\frac{\partial \mathbf{u}}{\partial t} + (\mathbf{u} \cdot \nabla)\left(\frac{\mathbf{u}}{\phi}\right) = -\frac{1}{\rho_f}\nabla(\phi p) + v_e \nabla^2 \mathbf{u} + \mathbf{F}, \qquad (2)$$

Temperature equation of fluid phase

$$\phi(\rho c_p)_f \frac{\partial T_f}{\partial t} + (\rho c_p)_f \mathbf{u} \cdot \nabla T_f = \phi \nabla \cdot (k_f \nabla T_f) + h_v (T_s - T_f) + \phi q_f''', \qquad (3)$$

Temperature equation of solid phase

$$(1-\phi)(\rho c_p)_s \frac{\partial T_s}{\partial t} = (1-\phi)\nabla \cdot (k_s \nabla T_s) + h_v (T_f - T_s) + (1-\phi) q_s''', \qquad (4)$$

where $\rho$ is the density, $\mathbf{u}$, $T$ and $p$ are the volume-averaged velocity, temperature and pressure, respectively, $\phi$ is the porosity, $v_e$ is the effective kinematic viscosity, $k$ is the thermal conductivity, $c_p$ is the specific heat at constant pressure, $q'''$ is the internal heat source (rate of volumetric heat generation), $h_v$ is the volumetric heat transfer coefficient, the subscript "$f$" and "$s$" denote the fluid and solid phases, respectively. **F** denotes the total body force induced by the porous media and other external force fields, which can be expressed as [39]

$$\mathbf{F} = -\frac{\phi v_f}{K}\mathbf{u} - \frac{\phi F_\phi}{\sqrt{K}}|\mathbf{u}|\mathbf{u} + \phi\mathbf{G},  \tag{5}$$

where $K$ is the permeability, $v_f$ is the kinematic viscosity of the fluid ($v_f$ is not necessarily the same as $v_e$). Based on Boussinesq approximation, the buoyancy force **G** induced by the gravitational field is given by

$$\mathbf{G} = -\mathbf{g}\beta(T_f - T_0),  \tag{6}$$

where **g** is the gravitational acceleration, $\beta$ is the thermal expansion coefficient, and $T_0$ is the reference temperature.

The inertial coefficient $F_\phi$ (Forchheimer coefficient) and the permeability $K$ depend on the geometry of the permeable media. They cannot be calculated analytically or measured directly. For flow over spherical particles, according to Ergun's experimental investigations [40], the inertial coefficient $F_\phi$ and the permeability $K$ can be expressed as [41]

$$F_\phi = \frac{1.75}{\sqrt{150\phi^3}}, \quad K = \frac{\phi^3 d_p^2}{150(1-\phi)^2},  \tag{7}$$

where $d_p$ is the diameter of the spherical particle.

Convection heat transfer in porous media under LTNE condition governed by Eqs. (1)-(4) is characterized by several dimensionless parameters: the Rayleigh number $Ra$, the Prandtl number $Pr$, the Darcy number $Da$, the Reynolds number $Re$, the viscosity ratio $J_e$, the solid-to-fluid thermal conductivity ratio $\gamma_{sf}$, the solid-to-fluid thermal diffusivity ratio $\Gamma_{sf}$, the solid-to-fluid heat capacity ratio $\sigma$, the dimensionless volumetric heat transfer coefficient $H_v$, the Fourier number $Fo$ (dimensionless time), which are defined as follows:

$$Ra = \frac{g\beta\Delta T L^3}{(\nu\alpha)_f}, \quad Pr = \frac{\nu_f}{\alpha_f}, \quad Da = \frac{K}{L^2}, \quad Re = \frac{Lu_c}{\nu_f}, \quad J_e = \frac{\nu_e}{\nu_f}, \quad \Gamma_{sf} = \frac{\alpha_s}{\alpha_f},$$

$$\gamma_{sf} = \frac{k_s}{k_f}, \quad \sigma = \frac{(\rho c_p)_s}{(\rho c_p)_f}, \quad H_\nu = \frac{h_\nu L^2}{k_f}, \quad Fo = \frac{tL^2}{\alpha_f}$$

where $L$ is the characteristic length, $\Delta T$ is the characteristic temperature, $u_c$ is the characteristic velocity, $\alpha_f = k_f/(\rho c_p)_f$ and $\alpha_s = k_s/(\rho c_p)_s$ are thermal diffusivities of the fluid and solid phases, respectively.

### 3. MRT-LB model for convection heat transfer in porous media under LTNE condition

The MRT model, as an important extension of the matrix LB method of Higuera et al. [9], was proposed by d'Humières in 1992 [42]. In the LB community, it has been widely accepted that the MRT model is superior over its BGK counterparts in terms of numerical stability and accuracy [43-47]. In recent years, several temperature-based MRT-LB models constructed within the framework of the double-distribution-function (DDF) approach have been used to study natural convection flows [48-51]. Numerical results indicate that the DDF-MRT method for incompressible thermal flows has a second-order convergence rate in space [49, 51]. In our previous studies [34], a DDF-MRT model has also been proposed to study convection heat transfer in porous media under LTE condition. In this work, a thermal MRT-LB model for convection heat transfer in porous media under LTNE condition will be constructed within the framework of the three-distribution-function (TDF) approach, which can be viewed as an extension to our previous studies.

#### 3.1 MRT-LB equation for velocity field

For two-dimensional (2D) problems considered in this work, the two-dimensional nine-velocity (D2Q9) lattice is employed. The nine discrete velocities $\{\mathbf{e}_i\}$ of the D2Q9 lattice are given by [11]

$$\mathbf{e}_i = \begin{cases} (0,0), & i = 0, \\ \left(\cos\left[(i-1)\pi/2\right], \sin\left[(i-1)\pi/2\right]\right)c, & i = 1 \sim 4, \\ \left(\cos\left[(2i-9)\pi/4\right], \sin\left[(2i-9)\pi/4\right]\right)\sqrt{2}c, & i = 5 \sim 8, \end{cases} \quad (8)$$

where $c = \delta_x/\delta_t$ is the lattice speed with $\delta_t$ and $\delta_x$ being the discrete time step and lattice spacing, respectively. The lattice speed $c$ is set to be 1 ($\delta_x = \delta_t$) in this work.

For velocity field, the MRT-LB equation with an forcing term can be written as [34, 46, 52]

$$f_i(\mathbf{x}+\mathbf{e}_i\delta_t, t+\delta_t) = f_i(\mathbf{x},t) - \tilde{\Lambda}_{ij}\left(f_j - f_j^{eq}\right)\Big|_{(\mathbf{x},t)} + \delta_t\left(S_i - 0.5\tilde{\Lambda}_{ij}S_j\right). \quad (9)$$

where $f_i(\mathbf{x},t)$ is the density distribution function with velocity $\mathbf{e}_i$ at position $\mathbf{x}$ and time $t$, $f_i^{eq}(\mathbf{x},t)$ is the equilibrium distribution function, $S_i$ is the forcing term, and $\tilde{\Lambda} = [\tilde{\Lambda}_{ij}]$ is the collision matrix.

MRT-LB equation (9) consists of two steps: the collision process and streaming process. Through a transformation matrix $\mathbf{M}$, the collision process can be executed in the moment space:

$$\mathbf{m}^*(\mathbf{x},t) = \mathbf{m}(\mathbf{x},t) - \Lambda\left(\mathbf{m} - \mathbf{m}^{eq}\right)\Big|_{(\mathbf{x},t)} + \delta_t\left(\mathbf{I} - \frac{\Lambda}{2}\right)\tilde{\mathbf{S}} \quad (10)$$

while the streaming process is still carried out in the velocity space:

$$f_i(\mathbf{x}+\mathbf{e}_i\delta_t, t+\delta_t) = f_i^*(\mathbf{x},t), \quad (11)$$

where $\Lambda = \mathbf{M}\tilde{\Lambda}\mathbf{M}^{-1} = \text{diag}(s_0, s_1, \ldots, s_8)$ is the relaxation matrix ($\{s_i\}$ are relaxation rates), $\mathbf{I}$ is the identity matrix. The bold-face symbols $\mathbf{m}$, $\mathbf{m}^{eq}$ and $\tilde{\mathbf{S}}$ denote 9-dimensional column vectors of moments as follows:

$$\mathbf{m} = \mathbf{M}\mathbf{f} = |m\rangle, \quad \mathbf{m}^{eq} = \mathbf{M}\mathbf{f}^{eq} = |m^{eq}\rangle, \quad \tilde{\mathbf{S}} = \mathbf{M}\mathbf{S} = |\tilde{S}\rangle, \quad (12)$$

where $\mathbf{f} = |f\rangle$, $\mathbf{f}^{eq} = |f^{eq}\rangle$, and $\mathbf{S} = |S\rangle$. For brevity, the Dirac notation $|\cdot\rangle$ is used to denote a 9-dimenaional column vector, e.g., $|m\rangle = (m_0, m_1, \ldots, m_8)^\mathrm{T}$. The post-collision distribution functions $\{f_i^*\}$ can be determined by $\mathbf{f}^* = |f^*\rangle = \mathbf{M}^{-1}\mathbf{m}^*$.

For the D2Q9 model, the transformation matrix $\mathbf{M}$ is given by ($c = 1$) [43]

$$\mathbf{M} = \begin{bmatrix} 1 & 1 & 1 & 1 & 1 & 1 & 1 & 1 & 1 \\ -4 & -1 & -1 & -1 & -1 & 2 & 2 & 2 & 2 \\ 4 & -2 & -2 & -2 & -2 & 1 & 1 & 1 & 1 \\ 0 & 1 & 0 & -1 & 0 & 1 & -1 & -1 & 1 \\ 0 & -2 & 0 & 2 & 0 & 1 & -1 & -1 & 1 \\ 0 & 0 & 1 & 0 & -1 & 1 & 1 & -1 & -1 \\ 0 & 0 & -2 & 0 & 2 & 1 & 1 & -1 & -1 \\ 0 & 1 & -1 & 1 & -1 & 0 & 0 & 0 & 0 \\ 0 & 0 & 0 & 0 & 0 & 1 & -1 & 1 & -1 \end{bmatrix}. \tag{13}$$

With the transformation matrix $\mathbf{M}$ given above, the moment vector $\mathbf{m}$ is defined as

$$\mathbf{m} = |m\rangle = \left( \rho, e, \varepsilon, j_x - \frac{\delta_t}{2}\rho F_x, q_x, j_y - \frac{\delta_t}{2}\rho F_y, q_y, p_{xx}, p_{xy} \right)^{\mathrm{T}}, \tag{14}$$

where $j_x = \rho u_x$ and $j_y = \rho u_y$ are $x$- and $y$-components of the momentum $\mathbf{j} = (j_x, j_y) = \rho \mathbf{u}$, respectively, $F_x$ and $F_y$ are $x$- and $y$-components of the total body force $\mathbf{F}$, respectively. The equilibrium moments $\{m_i^{eq}\}$ are given by [34]

$$m_0^{eq} = \rho, \ m_1^{eq} = -2\rho + \frac{3\rho |\mathbf{u}|^2}{\phi}, \ m_2^{eq} = \rho - \frac{3\rho |\mathbf{u}|^2}{\phi}, \ m_3^{eq} = \rho u_x, \ m_4^{eq} = -\rho u_x,$$

$$m_5^{eq} = \rho u_y, \ m_6^{eq} = -\rho u_y, \ m_7^{eq} = \frac{\rho(u_x^2 - u_y^2)}{\phi}, \ m_8^{eq} = \frac{\rho u_x u_y}{\phi}. \tag{15}$$

The components of the forcing term $\tilde{\mathbf{S}}$ in the moment space are given as follows [34]

$$\tilde{S}_0 = 0, \ \tilde{S}_1 = \rho F_x, \ \tilde{S}_2 = \rho F_y, \ \tilde{S}_3 = \frac{2\rho(u_x F_x + u_y F_y)}{\phi}, \ \tilde{S}_4 = \frac{2\rho(u_x F_x - u_y F_y)}{\phi},$$

$$\tilde{S}_5 = \frac{\rho(u_x F_y + u_y F_x)}{\phi}, \ \tilde{S}_6 = \frac{1}{3}\rho F_y, \ \tilde{S}_7 = \frac{1}{3}\rho F_x, \ \tilde{S}_8 = \frac{2}{3}\frac{\rho(u_x F_y + u_y F_x)}{\phi}. \tag{16}$$

The relaxation matrix $\Lambda$ is diagonal and is given by

$$\Lambda = \mathrm{diag}(s_0, s_1, s_2, s_3, s_4, s_5, s_6, s_7, s_8).$$

$$= \mathrm{diag}(1, s_e, s_\varepsilon, 1, s_q, 1, s_q, s_\nu, s_\nu). \tag{17}$$

The macroscopic fluid density $\rho$ and velocity $\mathbf{u}$ are defined as

$$\rho = \sum_{i=0}^{8} f_i, \tag{18}$$

$$\rho \mathbf{u} = \sum_{i=0}^{8} \mathbf{e}_i f_i + \frac{\delta_t}{2} \rho \mathbf{F}. \qquad (19)$$

The macroscopic fluid pressure $p$ is defined as $p = \rho c_s^2 / \phi$. Note that Eq. (19) is a nonlinear equation for the velocity $\mathbf{u}$. By introducing a temporal velocity $\mathbf{v}$, the macroscopic fluid velocity $\mathbf{u}$ can be calculated explicitly by

$$\mathbf{u} = \frac{\mathbf{v}}{l_0 + \sqrt{l_0^2 + l_1 |\mathbf{v}|}}, \qquad (20)$$

where

$$\mathbf{v} = \sum_{i=0}^{8} \mathbf{e}_i f_i / \rho + \frac{\delta_t}{2} \phi \mathbf{G}, \qquad (21)$$

$$l_0 = \frac{1}{2}\left(1 + \phi \frac{\delta_t}{2} \frac{v_f}{K}\right), \quad l_1 = \phi \frac{\delta_t}{2} \frac{F_\phi}{\sqrt{K}}. \qquad (22)$$

Through the Chapman-Enskog analysis of the MRT-LB equation (9), the generalized Navier-Stokes equations (1) and (2) can be recovered in the incompressible limit. The effective kinematic viscosity $v_e$ and the bulk viscosity $v_B$ are defined by

$$v_e = c_s^2 \left(\frac{1}{s_v} - \frac{1}{2}\right)\delta_t, \quad v_B = c_s^2 \left(\frac{1}{s_e} - \frac{1}{2}\right)\delta_t, \qquad (23)$$

respectively, where $s_{7,8} = s_v = 1/\tau_v$, and $c_s = c/\sqrt{3}$ is the sound speed.

The equilibrium distribution function $f_i^{eq}$ in the velocity space is given by [29]

$$f_i^{eq} = w_i \rho \left[1 + \frac{\mathbf{e}_i \cdot \mathbf{u}}{c_s^2} + \frac{(\mathbf{e}_i \cdot \mathbf{u})^2}{2\phi c_s^4} - \frac{|\mathbf{u}|^2}{2\phi c_s^2}\right]. \qquad (24)$$

where $w_0 = 4/9$, $w_i = 1/9$ for $i = 1 \sim 4$, $w_i = 1/36$ for $i = 5 \sim 8$.

*3.2 MRT-LB equations for temperature fields*

In this subsection, two temperature-based MRT-LB equations are proposed to solve the temperature fields of fluid and solid phases based on the passive scalar approach [53]. The two-dimensional five-velocity (D2Q5) lattice is employed, the five discrete velocities $\{\mathbf{e}_i | i = 0 \sim 4\}$

are given in Eq. (8). For clarity, the temperature equations (3) and (4) are rewritten as follow:

$$\frac{\partial T_f}{\partial t} + \nabla \cdot \left( \frac{\mathbf{u} T_f}{\phi} \right) = \nabla \cdot \left( \alpha_f \nabla T_f \right) + \frac{h_v \left( T_s - T_f \right)}{\phi \left( \rho c_p \right)_f} + \frac{q_f'''}{\left( \rho c_p \right)_f}, \tag{25}$$

$$\frac{\partial T_s}{\partial t} = \nabla \cdot \left( \alpha_s \nabla T_s \right) + \frac{h_v \left( T_f - T_s \right)}{(1-\phi) \left( \rho c_p \right)_s} + \frac{q_s'''}{\left( \rho c_p \right)_s}, \tag{26}$$

For the temperature fields governed by Eqs. (25) and (26), the temperature-based MRT-LB equations with source terms are given by [34]

$$\mathbf{g}_f \left( \mathbf{x} + \mathbf{e}_i \delta_t, t + \delta_t \right) = \mathbf{g}_f \left( \mathbf{x}, t \right) - \mathbf{N}^{-1} \mathbf{Q} \left( \mathbf{n}_f - \mathbf{n}_f^{eq} \right) \Big|_{(\mathbf{x}, t)} + \delta_t \mathbf{N}^{-1} \left( \mathbf{I} - \frac{\mathbf{Q}}{2} \right) \mathbf{\Psi}_f, \tag{27}$$

$$\mathbf{g}_s \left( \mathbf{x} + \mathbf{e}_i \delta_t, t + \delta_t \right) = \mathbf{g}_s \left( \mathbf{x}, t \right) - \mathbf{N}^{-1} \hat{\mathbf{Q}} \left( \mathbf{n}_s - \mathbf{n}_s^{eq} \right) \Big|_{(\mathbf{x}, t)} + \delta_t \mathbf{N}^{-1} \left( \mathbf{I} - \frac{\hat{\mathbf{Q}}}{2} \right) \mathbf{\Psi}_s, \tag{28}$$

where $\mathbf{n} = |n\rangle = \mathbf{Ng}$, $\mathbf{n}^{eq} = |n^{eq}\rangle = \mathbf{Ng}^{eq}$, $\mathbf{g} = |g\rangle$, $\mathbf{g}^{eq} = |g^{eq}\rangle$, $g_i(\mathbf{x}, t)$ and $g_i^{eq}(\mathbf{x}, t)$ are the temperature distribution function and the corresponding equilibrium distribution function, respectively; $\mathbf{N}$ is the transformation matrix, $\mathbf{Q}$ and $\hat{\mathbf{Q}}$ are relaxation matrices, $\mathbf{I}$ is the identity matrix; $\mathbf{\Psi}_f = |\Psi_f\rangle$ and $\mathbf{\Psi}_s = |\Psi_s\rangle$ are source terms defined by

$$\mathbf{\Psi}_f = \mathbf{N} \tilde{\mathbf{\Psi}}_f = \mathbf{N} |\tilde{\Psi}_f\rangle, \quad \mathbf{\Psi}_s = \mathbf{N} \tilde{\mathbf{\Psi}}_s = \mathbf{N} |\tilde{\Psi}_s\rangle, \tag{29}$$

in which

$$\tilde{\Psi}_{fi} = \tilde{w}_i Sr_f = \tilde{w}_i \left[ \frac{h_v \left( T_s - T_f \right)}{\phi \left( \rho c_p \right)_f} + \frac{q_f'''}{\left( \rho c_p \right)_f} \right], \tag{30}$$

$$\tilde{\Psi}_{si} = \tilde{w}_i Sr_s = \tilde{w}_i \left[ \frac{h_v \left( T_f - T_s \right)}{(1-\phi) \left( \rho c_p \right)_s} + \frac{q_s'''}{\left( \rho c_p \right)_s} \right], \tag{31}$$

where $\{\tilde{w}_i\}$ are weight coefficients given by $\tilde{w}_0 = 1 - \varpi$, $\tilde{w}_{1 \sim 4} = \varpi/4$ ($0 < \varpi < 1$).

Through the transformation matrix $\mathbf{N}$, the collision processes of the MRT-LB equations (27) and (28) can be executed in the moment space:

$$\mathbf{n}_f^* = \mathbf{n}_f - \mathbf{Q} \left( \mathbf{n}_f - \mathbf{n}_f^{eq} \right) + \delta_t \left( \mathbf{I} - \frac{\mathbf{Q}}{2} \right) \mathbf{\Psi}_f, \tag{32}$$

$$\mathbf{n}_s^* = \mathbf{n}_s - \hat{\mathbf{Q}}(\mathbf{n}_s - \mathbf{n}_s^{eq}) + \delta_t \left(\mathbf{I} - \frac{\hat{\mathbf{Q}}}{2}\right)\Psi_s, \tag{33}$$

The streaming processes are still carried out in the velocity space:

$$g_{fi}(\mathbf{x} + \mathbf{e}_i \delta_t, t + \delta_t) = g_{fi}^*(\mathbf{x}, t). \tag{34}$$

$$g_{si}(\mathbf{x} + \mathbf{e}_i \delta_t, t + \delta_t) = g_{si}^*(\mathbf{x}, t). \tag{35}$$

where $\mathbf{g}^* = |g^*\rangle = \mathbf{N}^{-1}\mathbf{n}^*$.

For the D2Q5 model, the transformation matrix $\mathbf{N}$ is given by ($c = 1$) [48, 49]

$$\mathbf{N} = \left[|1\rangle, |e_x\rangle, |e_y\rangle, |5\mathbf{e}^2 - 4\rangle, |e_x^2 - e_y^2\rangle\right]^{\mathrm{T}}$$

$$= \begin{bmatrix} 1 & 1 & 1 & 1 & 1 \\ 0 & 1 & 0 & -1 & 0 \\ 0 & 0 & 1 & 0 & -1 \\ -4 & 1 & 1 & 1 & 1 \\ 0 & 1 & -1 & 1 & -1 \end{bmatrix}. \tag{36}$$

The equilibrium moments $\{n_{fi}^{eq}\}$ and $\{n_{si}^{eq}\}$ are defined as

$$n_{f0}^{eq} = T_f, \quad n_{f1}^{eq} = \frac{u_x T_f}{\phi}, \quad n_{f2}^{eq} = \frac{u_y T_f}{\phi}, \quad n_{f3}^{eq} = (-4 + 5\varpi)T_f, \quad n_{f4}^{eq} = 0, \tag{37}$$

$$n_{s0}^{eq} = T_s, \quad n_{s1}^{eq} = 0, \quad n_{s2}^{eq} = 0, \quad n_{s3}^{eq} = (-4 + 5\varpi)T_s, \quad n_{s4}^{eq} = 0. \tag{38}$$

The relaxation matrices $\mathbf{Q}$ and $\hat{\mathbf{Q}}$ are given by

$$\mathbf{Q} = \mathrm{diag}(1, \zeta_\alpha, \zeta_\alpha, \zeta_e, \zeta_\nu), \tag{39}$$

$$\hat{\mathbf{Q}} = \mathrm{diag}(1, \eta_\alpha, \eta_\alpha, \eta_e, \eta_\nu), \tag{40}$$

The temperatures $T_f$ and $T_s$ can be calculated by

$$T_f = \sum_{i=0}^{4} g_{fi} + \frac{\delta_t}{2} Sr_f, \quad T_s = \sum_{i=0}^{4} g_{si} + \frac{\delta_t}{2} Sr_s \tag{41}$$

where

$$Sr_f = \frac{h_v(T_s - T_f)}{\phi(\rho c_p)_f} + \frac{q_f'''}{(\rho c_p)_f}, \quad Sr_s = \frac{h_v(T_f - T_s)}{(1-\phi)(\rho c_p)_s} + \frac{q_s'''}{(\rho c_p)_s}. \tag{42}$$

Note that the source terms also contain the temperatures $T_f$ and $T_s$. Obviously, $Sr_f$ and $Sr_s$ are in linear relationship with $T_f$ and $T_s$. Therefore, the temperatures $T_f$ and $T_s$ can be calculated

explicitly by

$$T_f = \frac{1}{1+d_0+d_1}\left[(1+d_1)\left(\sum_i g_{fi} + \frac{\delta_t}{2}Q_f'''\right) + d_0\left(\sum_i g_{si} + \frac{\delta_t}{2}Q_s'''\right)\right], \tag{43}$$

$$T_s = \frac{1}{1+d_0+d_1}\left[d_1\left(\sum_i g_{fi} + \frac{\delta_t}{2}Q_f'''\right) + (1+d_0)\left(\sum_i g_{si} + \frac{\delta_t}{2}Q_s'''\right)\right], \tag{44}$$

where

$$Q_f''' = \frac{q_f'''}{(\rho c_p)_f}, \quad Q_s''' = \frac{q_s'''}{(\rho c_p)_s}, \tag{45}$$

$$d_0 = \frac{\delta_t}{2}\frac{h_v}{\phi(\rho c_p)_f}, \quad d_1 = \frac{\delta_t}{2}\frac{h_v}{(1-\phi)(\rho c_p)_s}. \tag{46}$$

Through the Chapman-Enskog analysis of the MRT-LB equations (27) and (28), the following macroscopic equations can be obtained

$$\partial_t T_f + \nabla\cdot\left(\frac{\mathbf{u}T_f}{\phi}\right) = \nabla\cdot\left[\alpha_f \nabla T_f + \delta_t\left(\zeta_\alpha^{-1}-0.5\right)\epsilon\partial_{t_1}\left(\frac{\mathbf{u}T_f}{\phi}\right)\right] + Sr_f, \tag{47}$$

$$\frac{\partial T_s}{\partial t} = \nabla\cdot(\alpha_s \nabla T_s) + Sr_s, \tag{48}$$

where

$$\alpha_f = c_{sT}^2\left(\zeta_\alpha^{-1}-\frac{1}{2}\right)\delta_t, \quad \alpha_s = c_{sT}^2\left(\eta_\alpha^{-1}-\frac{1}{2}\right)\delta_t, \tag{49}$$

here, $\zeta_{1,2}=\zeta_\alpha=1/\tau_{Tf}$, $\eta_{1,2}=\eta_\alpha=1/\tau_{Ts}$, $c_{sT}^2 = c^2\varpi/2 = \varpi/2$ ($c_{sT}$ is the sound speed of the D2Q5 model). Eq. (47) contains an unwanted term $\nabla\cdot\left\{\delta_t\left(\zeta_\alpha^{-1}-0.5\right)\epsilon\partial_{t_1}\left(\mathbf{u}T_f/\phi\right)\right\}$ as compared with equation (25). According to Refs. [54, 55], a correction term $\delta_t\mathbf{N}^{-1}(\mathbf{I}-0.5\mathbf{Q})\mathbf{C}$ ($\mathbf{C}=\mathbf{N}\tilde{\mathbf{C}}=\mathbf{N}|\tilde{C}_i\rangle$ and $\tilde{C}_i = \tilde{w}_i T_f (\mathbf{e}_i\cdot\mathbf{F})/(\phi c_{sT}^2)$) should be added into the MRT-LB equation (27) to eliminate the effect of the unwanted term. For incompressible thermal flows considered in this work, the unwanted term can be neglected with little effect.

The equilibrium distribution functions $g_{fi}^{eq}$ and $g_{si}^{eq}$ in the velocity space are given by

$$g_{fi}^{eq} = \begin{cases} (1-\varpi)T_f, & i = 0 \\ \frac{1}{4}\varpi T_f + \frac{1}{2}\frac{(\mathbf{e}_i\cdot\mathbf{u})T_f}{\phi}, & i = 1\sim 4 \end{cases}, \tag{50}$$

$$g_{si}^{eq} = \tilde{w}_i T_s. \qquad (51)$$

## 4. Numerical simulations

In this section, numerical simulations of natural convection in a square cavity filled with a metallic porous medium are carried out to validate the present MRT-LB model. The schematic of the problem is shown in Fig. 1. The height of the cavity is $L$. The horizontal walls are adiabatic, while the left and right walls are kept at constant temperatures $T_h$ and $T_c$, respectively ($T_h > T_c$). In simulations, we set $\delta_t = \delta_x = \delta_y = 1$ ($c=1$), $\varpi = 2/5$ ($c_{sT}^2 = c^2/5 = 1/5$), $T_h = 11$, $T_c = T_0 = 1$, $\Delta T = T_h - T_c = 10$, $q_f''' = q_s''' = 0$, $s_e = s_\varepsilon = 1.1$, $s_q = 1.2$, $\zeta_e = \zeta_v = 1.1$, and $\eta_e = \eta_v = 1.1$. The characteristic parameters are set as follows: $J_e = 1$, $\sigma = 1$, $\gamma_{sf} = 1000$, $\phi = 0.8$, $F_\phi = 0.068$, $Pr = 1.0$, and $Da = 10^{-4}$.

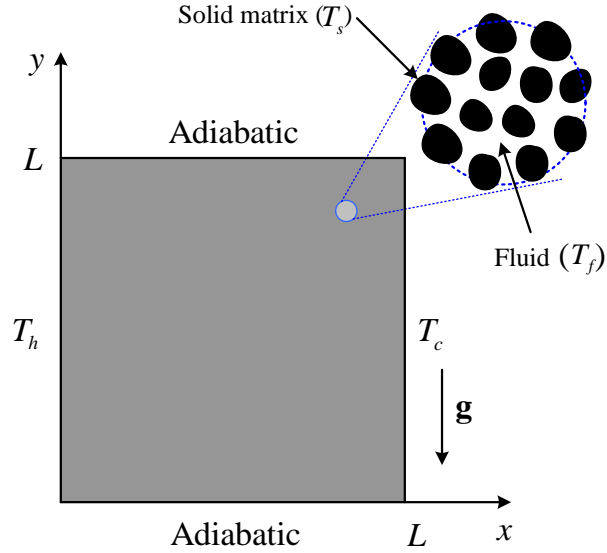

Fig. 1. Schematic of natural convection in a square cavity filled with a metallic porous medium.

Numerical simulations for $Ra = 10^6$ and $Ra = 10^8, 10^9$ are carried out based on $128 \times 128$ and $192 \times 192$ uniform meshes, respectively. The non-equilibrium extrapolation scheme [56] is employed to treat the velocity and temperature boundary conditions. The following correlation is employed to determine the volumetric heat transfer coefficient $h_v$ [7]:

$$H_p = \frac{h_v D_p^2}{k_f} = 0.376 (Re)^{0.644} (Pr)^{0.37}, \qquad (52)$$

where $D_p = 0.0135L$ is the average pore size, $Re$ is the local Reynolds number defined by $Re = u_m D_p / v_f$ ($u_m$ is the local average velocity). The dimensionless temperatures $\theta_s$ and $\theta_f$ are defined by $\theta_s = (T_s - T_0)/\Delta T$ and $\theta_f = (T_f - T_0)/\Delta T$, respectively.

Fig. 2(a) illustrates the variation of the solid-to-fluid temperature difference $\Delta \theta$ at the mid-height of the cavity ($y/L = 0.5$) at different times for $Ra = 10^6$ with $H_p = 0$ (without interphase heat exchange). For this case, the solid-to-fluid temperature difference $\Delta \theta$ depends on the relative response time of the two phases. Owing to the higher thermal diffusivity, the response time of the solid phase ($\sim \sigma/\gamma_{sf}$) is much faster than that of the fluid phase ($\sim O(1)$) and the solid phase is seen to reach the steady state much faster. Therefore, the solid-to-fluid temperature difference is very high at early times. The results for $H_p \neq 0$ ($H_p \neq 0$ denotes that $h_v$ is determined by Eq. (52)) are shown in Fig. 2(b). Clearly, the maximum solid-to-fluid temperature difference of $H_p \neq 0$ is less than that of $H_p = 0$ at time $Fo = 0.005$ due to interphase heat exchange. It is observed that the present results are in good agreement with the results reported in Ref. [7].

The solid temperature, fluid temperature and solid-to-fluid temperature difference at the mid-height of the cavity ($y/L = 0.5$) at the steady state for $Ra = 10^6, 10^8, 10^9$ with $H_p \neq 0$ are shown in Figs. 3 and 4. It can be seen that the present results are in excellent agreement with those reported in previous studies [7]. Moreover, Fig. 4 shows that the solid-to-fluid temperature difference increases with the increasing of $Ra$ (for $Ra = 10^9$, a maximum difference of about 10% can be seen), and therefore, the assumption of LTE must be discarded for large Rayleigh numbers. Fig. 5 shows the streamlines and isotherms of fluid and solid phases at the steady state for $Ra = 10^6, 10^8, 10^9$ with $H_p \neq 0$. Fig. 5 clearly shows that the solid-to-fluid temperature difference increases with the increasing of $Ra$, which is in accordance with those observed from Fig. 4.

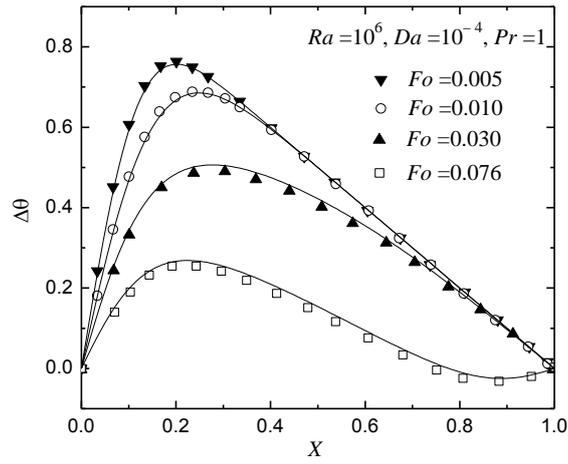

(a) $H_p = 0$

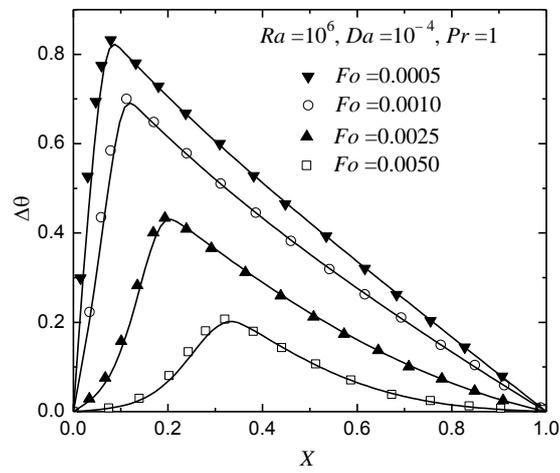

(b) $H_p \neq 0$

Fig. 2. Solid-to-fluid temperature difference at the mid-height of the cavity ($y/L = 0.5$) at different times for $Ra = 10^6$ (lines: MRT-LB results; symbols: results in Ref. [7]).

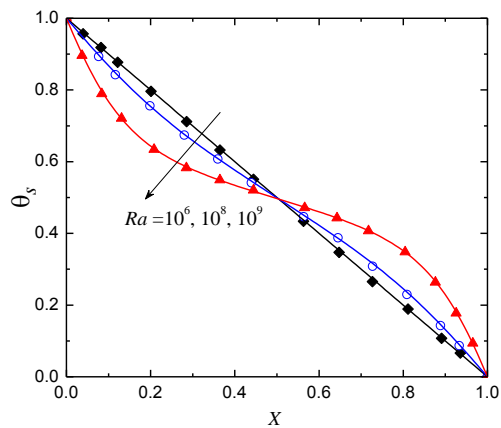

(a) Solid temperatures

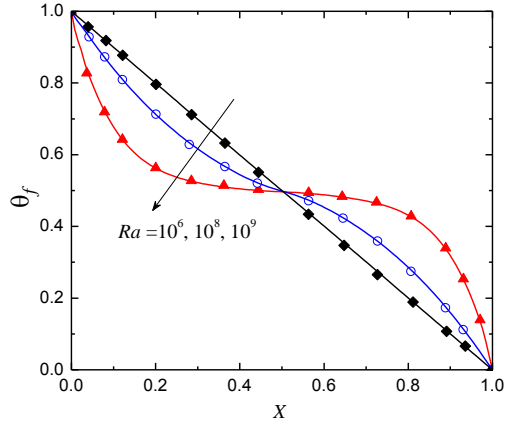

(b) Fluid temperatures

Fig. 3. Solid (a) and fluid (b) temperatures at the mid-height of the cavity ($y/L = 0.5$) at the steady state for $Ra = 10^6, 10^8, 10^9$ with $H_p \neq 0$ (lines: MRT-LB results; symbols: results in Ref. [7]).

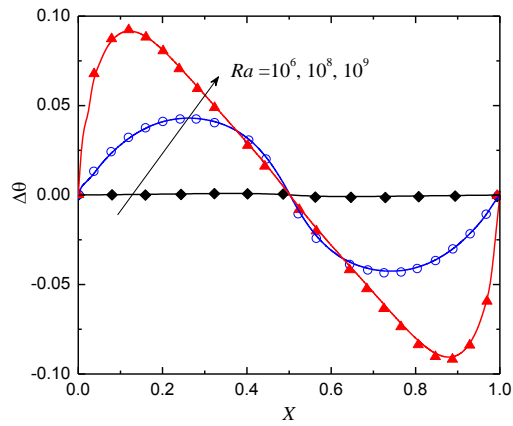

Fig. 4. Solid-to-fluid temperature difference at the mid-height of the cavity ($y/L = 0.5$) at the steady state for $Ra = 10^6, 10^8, 10^9$ with $H_p \neq 0$ (lines: MRT-LB results; symbols: results in Ref. [7]).

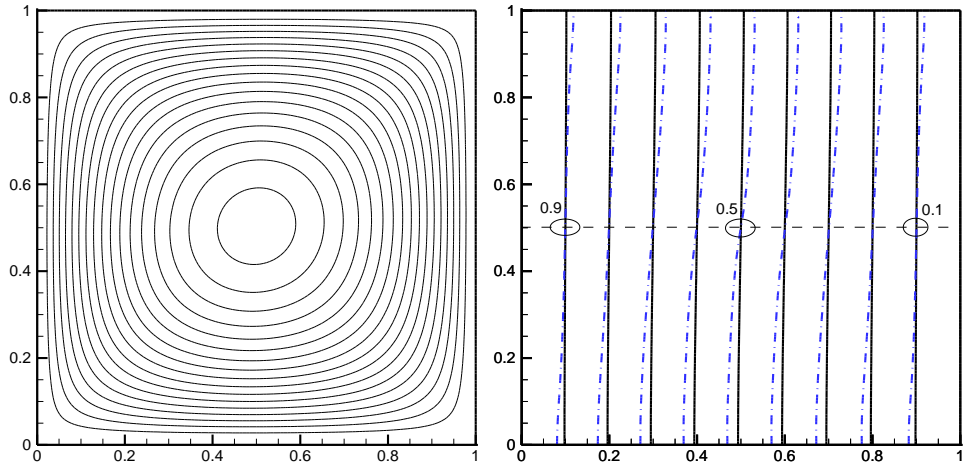

(a) $Ra = 10^6$

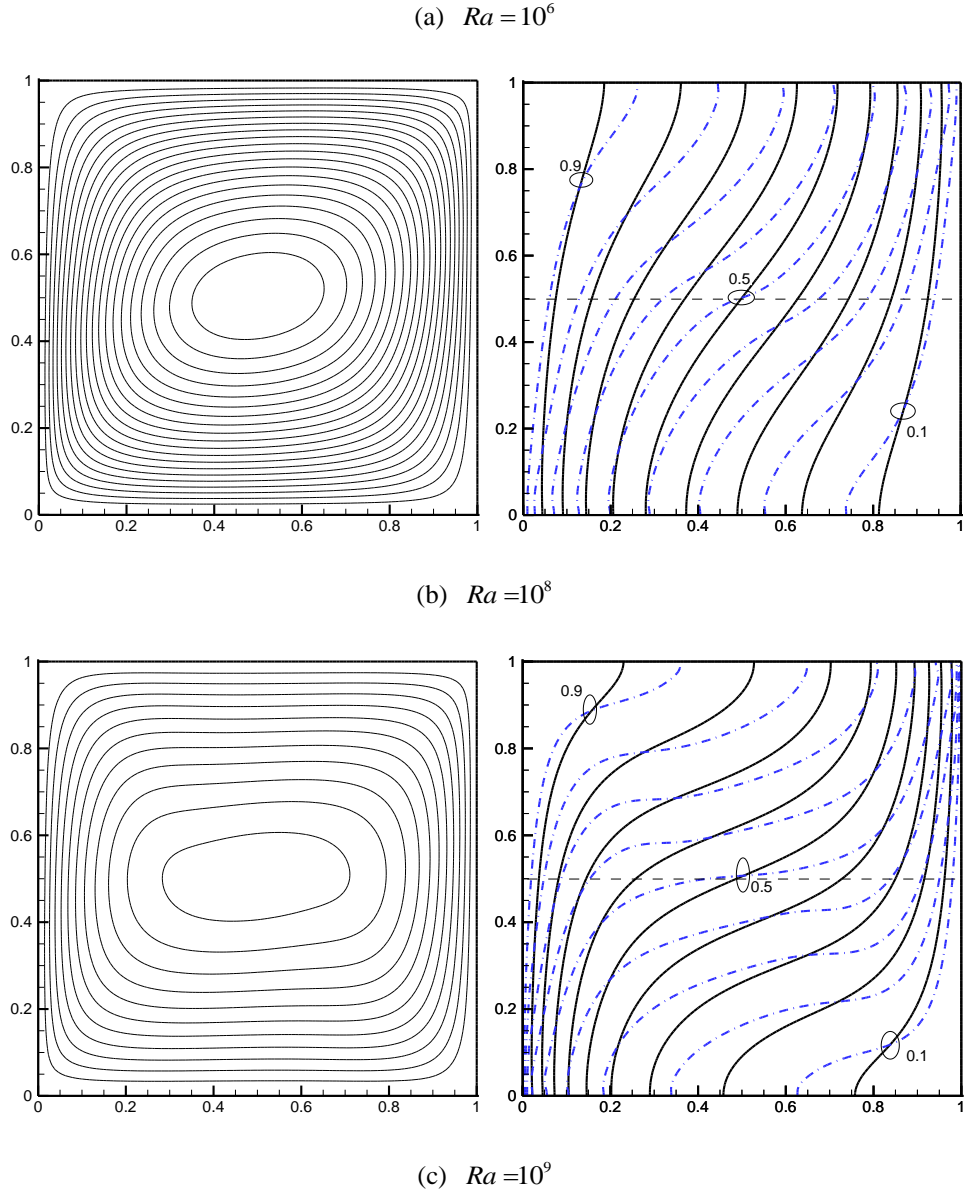

(b) $Ra = 10^8$

(c) $Ra = 10^9$

Fig. 5. Streamlines (left) and isotherms (right: fluid phase (red lines) and solid phase (blue dash-dot lines)) at the steady state for $Ra = 10^6, 10^8, 10^9$ with $H_p \neq 0$.

## 5. Summary

In this paper, a thermal MRT-LB model is proposed for convection heat transfer in porous media under LTNE condition. The model is constructed within the framework of the TDF approach: two temperature-based MRT-LB equations are proposed for the temperature fields of fluid and solid phases in addition to the MRT-LB equation of a density distribution function for the velocity field described by the generalized non-Darcy model. Moreover, the source terms accounting for the thermal non-equilibrium effects are simple and the model retains the inherent features of the standard LB method. Numerical simulations of natural convection in a square cavity filled with a metallic porous medium are carried out to validate the present model. It is found that the present numerical results are in good agreement with the numerical results reported in the literature.